\documentclass[11pt]{article}
\usepackage{graphicx}

\begin{document}

\title{
\textbf{Studies of critical phenomena and phase transitions in large lattices with Monte-Carlo
based non-perturbative approaches}}%
\author{J. Kaupu\v{z}s$^{1,2}$ \thanks{E--mail: \texttt{kaupuzs@latnet.lv}} \hspace*{1ex}, 
J. Rim\v{s}\={a}n$s^{1,2}$ \thanks{E--mail: \texttt{rimshans@latnet.lv}} \hspace*{1ex}, 
R. V. N. Melnik$^3$ \thanks{E--mail: \texttt{rmelnik@wlu.ca}} 
\\
$^1$ Institute of Mathematics and Computer Science, 
University of Latvia \\ 
29 Rai\c{n}a Blvd, LV--1459 Riga, Latvia \\ 
$^2$ Institute of Mathematical Sciences and Information Technologies, \\
University of Liepaja \\
14 Liela Street, Liepaja LV-3401, Latvia \\
$^3$ Wilfrid Laurier University \\
75 University Ave W, Waterloo, Ontario, Canada, N2L 3C5}

\date{\today}

\maketitle

\begin{abstract}
Critical phenomena and Goldstone mode effects in spin models with $O(n)$ rotational symmetry are 
considered. Starting with the Goldstone mode singularities in the $XY$ and $O(4)$ models,
we briefly review different theoretical concepts as well as state-of-the art Monte Carlo simulation results.
They support recent results of the GFD (grouping of Feynman diagrams) theory, 
stating that these singularities are described by certain
nontrivial exponents, which differ from those predicted earlier by perturbative treatments. 
Furthermore, we present the recent Monte Carlo simulation results of the three-dimensional Ising
model for very large lattices with linear sizes up to $L=1536$. These results are obtained, using
a parallel OpenMP implementation of the Wolff single cluster algorithm. The finite--size scaling 
analysis of the critical exponent $\eta$, assuming the usually accepted
correction-to-scaling exponent $\omega \approx 0.8$, shows that $\eta$ is likely to be somewhat larger
than the value $0.0335 \pm 0.0025$ of the perturbative 
renormalization group (RG) theory. Moreover, we have found that the 
actual data can be well described by different critical exponents: $\eta=\omega=1/8$
and $\nu=2/3$, found within the GFD theory.
\end{abstract}

\textbf{PACS} \hspace*{0.5ex} 05.10.Ln; 75.10.Hk; 05.50.+q

\section{Introduction}

Critical phenomena are observed in vicinity of phase transition points
in a variety of systems (e.~g., solids -- ferromagnets, ferroelectrics; 
fluids -- superfluid $\lambda$-transition; liquid--gas transition, etc.),
which manifest themselves in power--like singularities of physical observables,
described by critical exponents.
In many systems, like isotropic ferromagnets (e.~g., polycrystalline Ni), 
certain singularity is observed not only at the critical temperature $T=T_c$, 
but also at vanishing external field
$h \to 0$ for any temperature $T<T_c$. This phenomenon, known as the
Goldstone mode singularity, is also described by power--like divergences
and certain exponents. In this sense, it can be viewed as a critical phenomenon,
which takes place in vicinity of the critical line $h=0$ in the $T$--$h$ plane.

The vicinity of critical point (or line) is not the natural domain
of validity of any perturbation theory, therefore one should resort
the non--perturbative methods such as
\begin{itemize}
\item
Exact and rigorous analytical solution methods (transfer matrix methods, 
combinatorial methods, Bethe-ansatz),
\item
Conformal field theory analysis,
\item
Non--perturbative renormalization group (RG) analysis,
\item
Numerical transfer--matrix calculations,
\item
Molecular dynamics simulations,
\item
Monte Carlo (MC) simulations.
\end{itemize}
The exact solutions~\cite{Onsager,McCoyWu,Baxter} and conformal field analysis~\cite{FMS} are
powerful tools to determine the critical exponents
of two--dimensional models. However, these approaches are not helpful
in most of the three--dimensional cases. The MC method can be used here.
The exact non--perturbative RG equations are known for
various models on fractal lattices (see, e.~g.~\cite{ReMa} and references therein), 
and also in four dimensions a rigorous RG analysis has been made (see e.~g.,~\cite{Hara}). 
In most of other cases, the non--perturbative RG equations provide only approximations,
the accuracy of which cannot be well controlled (see, e.~g.,~\cite{Report1,Report2,KnonpRG}).
To the contrary, the MC method allows, in principle, to reach any desired accuracy.
Besides, it allows to treat much larger systems than the other numerical methods.
It is extremely important in studying critical phenomena.
Therefore, the MC method has no real alternative in non--perturbative determination
of critical exponents of the three--dimensional systems
like lattice spin ($O(n)$) models.

The general framework of perturbative methods includes the study
of critical point phenomena at $T \to T_c$, as well as the Goldstone
mode effects at $T<T_c$. Our aim is to verify nonperturbatively the 
validity of different theoretical perturbative
approaches. Hence, the MC test of the Goldstone mode singularities is one
of the important points here. 
We start with the Goldstone mode effects, since this case is 
simpler in the sense that it does not require the determination
of the critical temperature $T_c$, i.~e., there are less ambiguities
as compared to the $T \to T_c$ case.

\section{Goldstone mode singularities in the $O(n)$ models}
\label{sec:Goldstone}

Here we consider such lattice spin models, where the spin is an
$n$--component vector with $n \ge 2$. These are called $O(n)$ models due to the $O(n)$
global rotational symmetry exhibited by the $n$--vector  model 
in absence of the external field. The Hamiltonian $\mathcal{H}$ of such a model reads
\begin{equation}
\frac{\mathcal{H}}{T}=-\beta \left( \sum\limits_{\langle i j \rangle}
{\bf s}_i {\bf s}_j + \sum_i {\bf h s_i} \right) \;,
\end{equation} 
where $T$ is temperature, measured in energy units, 
${\bf s}_i$ is the spin variable ($n$--component vector of unit 
length) of the $i$th lattice site, $\beta$ is the 
coupling constant, and ${\bf h}$ is the external field with magnitude $\mid {\bf h} \mid=h$. 
The summation takes place over all pairs $\langle i j \rangle$ of the nearest neighbors in the lattice.

In the thermodynamic limit $L \to \infty$ below the critical point, i.~e., at $\beta>\beta_c$, the magnetization $M(h)$ and the 
Fourier--transformed transverse ($G_{\perp}({\bf k})$) and longitudinal ($G_{\parallel}({\bf k})$)
two--point correlation functions exhibit Goldstone mode power--law singularities:
\begin{eqnarray}
&&M(h) - M(+0) \propto h^{\rho} \quad \mbox{at} \quad h \to 0 \;, \\ 
&&G_{\perp}({\bf k}) \propto k^{-\lambda_{\perp}} \hspace{5ex} \mbox{at} \quad h=+0 \quad \mbox{and} \quad k \to 0 \;,\\
&&G_{\parallel}({\bf k}) \propto k^{-\lambda_{\parallel}} \hspace{6ex} \mbox{at} \quad h=+0 \quad \mbox{and} \quad k \to 0 \;.
\end{eqnarray}
According to the standard theory~\cite{Law2,HL,Tu,SH78,ABDS99}, $\lambda_{\perp} = 2$ and $\lambda_{\parallel}=4-d$ 
hold for $2<d<4$, and $\rho = 1/2$ is true in three dimensions.
More nontrivial universal values are expected according to~\cite{K2010}, such that
\begin{eqnarray}
&&d/2 < \lambda_{\perp} < 2 \;, \label{eq:pred1} \\
&&\lambda_{\parallel} = 2 \lambda_{\perp} - d \;, \label{eq:pred2} \\
&&\rho = (d/\lambda_{\perp})-1  \label{eq:pred3}
\end{eqnarray} 
hold for $2<d<4$. These relations have been obtained in~\cite{K2010} by analyzing self-consistent
diagram equations for the correlation functions without cutting the perturbation series.
This approach is based on certain grouping of Feynman diagrams introduced in~\cite{K_Ann01}, and
therefore is called the GFD theory. 

MC simulations have been performed~\cite{KMR07,KMR08,KMR10} to test these relations
for the $O(2)$ and $O(4)$ models on the simple cubic lattice.
The theoretical predictions~(\ref{eq:pred1}) and~(\ref{eq:pred2}) are confirmed by MC
simulation results in the 3D $O(4)$ model~\cite{KMR10}, where an estimate $\lambda_{\perp}=1.955 \pm 0.020$ has been found
for the transverse correlation function.
Moreover, it has been stated that the behavior of the longitudinal correlation function
is well consistent with $\lambda_{\parallel}$ being about $0.9$, in agreement with~(\ref{eq:pred2}) at
$\lambda_{\perp}$ being about $1.95$, but it is not consistent with the standard--theoretical prediction of $\lambda_{\parallel}=1$.
According to~(\ref{eq:pred3}), we have $1/2 < \rho <1$ in three dimensions. A reasonable numerical evidence
for this relation has been obtained in~\cite{KMR08} from the susceptibility data, where the MC estimate $\rho = 0.555(17)$
has been reported for the 3D $XY$ ($O(2)$) model. It corresponds to $\lambda_{\perp} = 1.929(21)$ according to~(\ref{eq:pred3}).

\section{Critical point singularities in the $n$--vector models}
\label{cps}

In vicinity of the phase transition point, various quantities have often
power--law singularities, which are described by the critical exponents.
For three--dimensional systems, exact results are difficult to obtain,
and one usually relies on approximate methods.
A review of  standard perturbative RG methods applied here can be
found, e.~g., in~\cite{Amit,Ma,Justin,Kleinert,PV}. 
The results of the Borel--resummation of the perturbation series for the critical
exponents, obtained within this approach, 
are reported in~\cite{GJ98}. We will further focus on the Monte Carlo testing of the 
theoretical predictions for the 3D Ising model, which is a particular case of $n=1$.
The critical exponent $\eta$, describing the $\sim k^{-2+\eta}$
singularity of the critical correlation function $G^*({\bf k})$, is of particular interest here.
According to~\cite{GJ98}, the most accurate theoretical value $\eta = 0.0335 \pm 0.0025$ is 
obtained from the series at fixed dimension $d=3$. The results of the resummation of
the $\epsilon$--expansion in~\cite{GJ98} are $\eta=0.036 \pm 0.005$ and $\eta=0.0365 \pm 0.0050$.
If all these estimates are correct within the error bars, then we have $0.0315< \eta < 0.036$.
This estimation fairly well agrees with the value $\eta=0.0366(8)$ extracted from the finite--size 
scaling analysis of the MC data within $L \in [10,128]$, reported in~\cite{HasRev}.
We will present the MC results for much larger lattices up to $L=1536$ to test the 
agreement more precisely. The other relevant here exponents are the correlation length
exponent $\nu$ and correction--to--scaling exponent $\omega$. The widely accepted values
for the 3D Ising model are $\nu \simeq 0.63$ and $\omega \simeq 0.8$. They are in agreement, e.~g.,
with the estimates $\nu = 0.6304 \pm 0.0013$ and $\omega = 0.799 \pm 0.011$ reported in~\cite{GJ98}.

As discussed in Sec.~\ref{sec:Goldstone}, the alternative theoretical approach, called the GFD theory,
provides promising results concerning the Goldstone mode singularities. Therefore, it is interesting
to verify the predictions of this theory also for the critical point singularities. 
In~\cite{K_Ann01}, a set of possible values of the
exact critical exponents for the $\varphi^4$ model in two ($d=2$) and three ($d=3$)
dimensions has been proposed:
\begin{equation} \label{eq:result}
\gamma = \frac{d+2j+4m}{d(1+m+j)-2j} \;; \hspace*{3ex}
\nu = \frac{2(1+m)+j}{d(1+m+j)-2j} \; ,
\end{equation}
where $\gamma$ is the susceptibility exponent, 
$m$ may have a natural value starting with 1, and $j$ is an integer equal or larger than $-m$.
Other critical exponents can be calculated from these ones, using the known scaling relations.
These values agree with the known exact solutions of the two--dimensional Ising model
($m=3$, $j=0$) and of the spherical model ($j/m \to \infty$). A prediction has been made also
for the 3D Ising model, i.~e., $\gamma=5/4$ and $\nu=2/3$, 
corresponding to $m=3$ and $j=0$, as in the two--dimensional case.
This value of $\nu$ is consistent with the logarithmic singularity of specific
heat (according to $\alpha=2-d \nu =0$) proposed earlier by Tseskis~\cite{Tseskis}. 
The exponents $\gamma=5/4$ and $\nu=2/3$ have been later conjectured for the 
3D Ising model by Zhang~\cite{Zhang}. The critical exponents
$\gamma = 9/8$ and $\nu=5/8$, calculated for the liquid--gas system by 
Bondarev~\cite{Bondarev,Bondarev10}, also
correspond to~(\ref{eq:result}). In this case $m=2$ and $j=-1$ hold.
As explained in~\cite{K_Ann01}, the equations~(\ref{eq:result}) are meaningful
for positive integer $n$, and we can have $j=j(n)$ and $m=m(n)$ in the case where 
the order parameter is an $n$--component vector.
The spatial dimensionality $d$ might be considered as a continuous
parameter in~(\ref{eq:result}) within $2 \le d \le 4$. At $n=1$,
the second--order phase transition at a finite temperature can be expected
also for $d<2$. However, according to the discussion in~\cite{K_surf05}, an analytic extension
of~(\ref{eq:result}) to this region, probably, is only formal and does not correspond
to any real system (fractal lattice).

According to~\cite{K_Ann01}, corrections to scaling can be represented by an expansion 
of correction factor (amplitude) in integer powers of $t^{2 \nu - \gamma}$ and 
$t^{2 \gamma - d \nu}$ at $t \to 0$, 
where $t$ is the reduced temperature. This expansion is simplified, since $(2 \gamma - d \nu)/(2 \nu - \gamma)$
is an integer number according to~(\ref{eq:result}), and $2 - \gamma/\nu = \eta$ holds according
to the known scaling relation. Hence, we obtain the expansion in powers of $t^{\theta}$,
where $\theta = \eta \nu$. 
In other words, the correction--to--scaling exponent in the expansions at $t \to 0$
is $\theta= \eta \nu$, and the corresponding correction--to--scaling exponent in the
finite--size--scaling analysis is $\omega= \theta/\nu = \eta$, if the first expansion coefficient
is nonzero. Allowing that some of the expansion coefficients are zero, we can have 
$\theta= \ell \eta \nu$, where $\ell$ is a positive integer. However,
our actual MC data support the most natural choice of $\ell=1$ for the 3D Ising model.
It implies that the expansion coefficients do not vanish in general,
but the nontrivial correction terms vanish only in some special cases, 
as the 2D Ising model or the spherical model, where $\omega=1$ and/or $\theta=1$. 
According to the numerical transfer matrix calculations in~\cite{trfm}, a nontrivial correction to 
finite--size scaling  with the exponent $\omega=\eta$, probably, 
exists in the two--point correlation function even in the 2D Ising model, 
although its amplitude is very small here.

The discussed here GFD critical exponents $\eta=1/8$ and $\nu=2/3$ for $n=1$ seem to be quite incompatible
with the MC data of the 3D Ising model, if we assume the usual correction--to--scaling exponent 
$\omega \simeq 0.8$. However, it turns out that our data for large enough sizes
are very well consistent even with the set of exponents $\eta=\omega=1/8$ and $\nu=2/3$. 
The disagreement of these exponents with those of the perturbative RG method
can also be understood based on a critical analysis~\cite{K_eprint}.

\section{Monte Carlo simulation results for the 3D Ising model}

We have simulated the 3D Ising model on a simple cubic lattice (at $h=0$), using 
the iterative method introduced in~\cite{KMR_2010} to adjust the coupling $\beta$ to the
pseudo-critical coupling $\widetilde{\beta}_c$, which corresponds to certain value $1.6$ of 
the quantity $U=\langle m^4 \rangle / \langle m^2 \rangle^2$, where $m$ is the magnetization per spin. 
The pseudo-critical coupling $\widetilde{\beta}_c$ tends to the true critical coupling $\beta_c$ at $L \to \infty$. 
By this method, we estimate the moments of energy $\varepsilon$ and magnetization $m$ per spin, i.~e., 
$\langle \varepsilon^k m^l \rangle$ for a set of $\beta$ values fluctuating around $\widetilde{\beta}_c$,
and then calculate these mean values and related quantities at any given $\beta$ near $\widetilde{\beta}_c$ 
by using the Taylor series expansion for $\ln \langle (-\varepsilon)^k m^l \rangle$. 
We have evaluated $\widetilde{\beta}_c(L)$
for various lattice sizes $L$ and then estimated $\beta_c$ by fitting these $\widetilde{\beta}_c(L)$ data.
Besides, we have calculated the susceptibility $\chi=L^3 \langle m^2 \rangle$ and the derivative $\partial Q /\partial \beta$,
where $Q =1/U$, at $\beta = \widetilde{\beta}_c(L)$, as well as at certain estimated critical couplings $\beta_c$,
in order to make a finite--size scaling analysis of the critical exponents.

This method has been tested and discussed in detail in~\cite{KMR_2010}, giving also some data 
for $L \le 1024$. Here we have extended the simulation results up to $L=1536$.
A parallel (OpenMP) implementation of the Wolff single cluster algorithm has been used in our simulations,
combining it with a sophisticated shuffling scheme for generation of a high--quality pseudo--random numbers,
as described in~~\cite{KMR_2010}. Several tests have been made in~\cite{KMR_2010} to verify the
quality of the pseudo--random numbers. Here we have tested and used also the lagged Fibonacci 
pseudo--random number generator (PRNG) with multiplication operation and the lags $r=24$, $s=55$ 
(see~\cite{MC}). We have improved it with the standard shuffling scheme~\cite{MC}, using the shuffling box
of the size $10\,000$ in this case. Although the lagged Fibonacci generator with additive operation produces
certain correlations in the sequence of pseudo--random numbers, which can be well detected by the
directed--random--walk test~\cite{SB}, this defect is practically not observed in the actual multiplicative case.
Like in~\cite{KMR_2010}, we have verified it by performing such test with $10^{12}$ trajectories.
Good results have been obtained in this test both for the original PRNG and for the one improved by the shuffling.
It indicates that the multiplicative lagged Fibonacci generator works fine in applications with cluster algorithms,
as it also has been mentioned in~\cite{MB04}.

The simulations of the two largest lattices with $L=1280$ and $L=1536$ have been performed with
the shuffling scheme of~\cite{KMR_2010} and partly (24 usable iterations from 56 or 54 ones
at $L=1280$ and $L=1536$, respectively) also with the improved lagged Fibonacci generator.
For complete confidence, we have verified that
the simulation results of the two generators agree within the statistical error bars,
whereas the final simulated values have been obtained by summarizing the data from both of them.
Note that one iteration included $5\,280\,000$ MC (Wolff cluster algorithm) steps at $L=1280$
and $6\,720\,000$ MC steps at $L=1536$, the MC measurements being performed after
each $160$ and $192$ steps, respectively.

Our simulation results for the pseudo-critical coupling $\widetilde{\beta}_c$, as well as for
the corresponding values of $\chi/L^2$ and $\partial Q /\partial \beta$ depending on the lattice size $L$
are given in Tab.~\ref{tab1}. In Tab.~\ref{tab2}, the latter two quantities are given, calculated at 
$\beta=0.221654604$ and $\beta=0.221654615$, corresponding to two different estimates of the critical
coupling. Besides, we have calculated these quantities also at $\beta=0.2216545$ for $L \le 128$ and have verified
that the values agree within the error bars with those reported for this $\beta$ in~\cite{HasRev}.
\begin{table}
\caption{The values of $\widetilde \beta_c$, as well as $\chi/L^2$ and 
$10^{-3} \partial Q /\partial \beta$ at $\beta = \widetilde{\beta}_c$ depending on $L$.}
\label{tab1}
\begin{center}
\begin{tabular}{|c|c|c|c|}
\hline
\rule[-2mm]{0mm}{7mm} 
L & $\widetilde \beta_c$ & $\chi/L^2$  & $10^{-3} \partial Q /\partial \beta$  \\
\hline
1536 & 0.2216546081(114) & 1.1900(22) & 96.01(73)  \\
1280 & 0.2216546524(136) & 1.2009(25) & 72.04(53)  \\
1024 & 0.221654625(22)   & 1.2046(28) & 50.57(45)  \\
864  & 0.221654635(25)   & 1.2165(21) & 38.71(25)  \\
768  & 0.221654672(27)   & 1.2212(20) & 31.89(20)  \\
640  & 0.221654615(31)   & 1.2281(17) & 23.95(12)  \\
512  & 0.221654662(45)  & 1.2367(16)  & 16.785(77)  \\
432  & 0.221654637(58)  & 1.2450(18)  & 12.907(59)  \\
384  & 0.221654567(65)  & 1.2480(16)  & 10.627(50)  \\
320  & 0.221654716(75)  & 1.2578(16)  & 7.967(36)  \\
256  & 0.22165460(11)  & 1.2656(15)   & 5.577(24)  \\
216  & 0.22165460(13)  & 1.2726(12)   & 4.288(14)  \\
192  & 0.22165425(16)  & 1.2734(14)   & 3.533(14)  \\
160  & 0.22165414(18)  & 1.2818(11)   & 2.6495(87) \\
128  & 0.22165430(20)  & 1.2913(10)   & 1.8643(49) \\
108  & 0.22165376(26)  & 1.2969(10)   & 1.4170(36) \\
96   & 0.22165369(32)  & 1.3012(10)   & 1.1796(28) \\
80   & 0.22165278(32)  & 1.30659(74)  & 0.8822(18) \\
64   & 0.22165159(52)  & 1.31466(78)  & 0.6192(11) \\
54   & 0.22164968(56)  & 1.31916(76)  & 0.47334(81) \\
48   & 0.22164790(69)  & 1.32164(66)  & 0.39331(63) \\
40   & 0.22164383(80)  & 1.32562(64)  & 0.29424(40) \\
32   & 0.22163444(98)  & 1.32835(59)  & 0.20586(25) \\
27   & 0.2216212(11)   & 1.32829(52)  & 0.15703(16) \\
24   & 0.2216125(12)   & 1.33027(47)  & 0.13076(13) \\
20   & 0.2215821(17)   & 1.32860(42)  & 0.097717(87) \\
16   & 0.2215235(18)   & 1.32510(34)  & 0.068538(48) \\
\hline
\end{tabular}
\end{center}
\end{table}
\begin{table}
\caption{The values of  $\chi/L^2$ and 
$10^{-3} \partial Q /\partial \beta$ at $\beta=0.221654604$ and $\beta=0.221654615$
depending on $L$.}
\label{tab2}
\begin{center}
\begin{tabular}{|c|c|c|c|c|}
\hline
& \multicolumn{2}{|c|}{\rule[-2.5mm]{0mm}{7mm} $\beta=0.221654604$} 
& \multicolumn{2}{|c|}{\rule[-2.5mm]{0mm}{7mm} $\beta=0.221654615$} 
\\ \cline{2-5}
\raisebox{1.5ex}{L} 
& \rule[-2mm]{0mm}{4mm}
$\chi/L^2$  & $10^{-3}\partial Q /\partial \beta$ & $\chi/L^2$ & $10^{-3}\partial Q /\partial \beta$ \\ \hline
1536 & 1.1882(38) & 95.94(69) & 1.1930(38) & 96.12(69) \\
1280 & 1.1849(30) & 71.62(45) & 1.1885(30) & 71.72(45) \\
1024 & 1.1998(37) & 50.48(40) & 1.2023(37) & 50.53(40) \\
864  & 1.2110(32) & 38.63(22) & 1.2130(32) & 38.66(22) \\
768  & 1.2110(31) & 31.77(18) & 1.2126(32) & 31.79(18) \\
640  & 1.2268(27) & 23.94(10) & 1.2280(27) & 23.95(10) \\
512  & 1.2321(27) & 16.756(69) & 1.2330(27) & 16.762(69) \\
432  & 1.2429(24) & 12.898(52) & 1.2436(24) & 12.901(52) \\
384  & 1.2499(24) & 10.634(44) & 1.2505(24) & 10.636(44) \\
320  & 1.2534(22) & 7.955(33) & 1.2538(22) & 7.956(33) \\
256  & 1.2656(23) & 5.577(22) & 1.2659(23) & 5.577(22) \\
216  & 1.2727(21) & 4.288(13) & 1.2729(21) & 4.288(13) \\
192  & 1.2795(21) & 3.540(12) & 1.2797(21) & 3.541(12) \\
160  & 1.2879(17) & 2.6550(78) & 1.2880(17) & 2.6551(78) \\
128  & 1.2941(13) & 1.8660(42) & 1.2942(13) & 1.8661(42) \\
108  & 1.3028(13) & 1.4198(32) & 1.3029(13) & 1.4198(32) \\
96   & 1.3066(14) & 1.1817(25) & 1.3066(14) & 1.1817(25) \\
80   & 1.3147(11) & 0.8845(16) & 1.3147(11) & 0.8845(16) \\
64   & 1.3241(12) & 0.62115(96) & 1.3241(12) & 0.62116(96) \\
54   & 1.33098(85) & 0.47518(69) & 1.33101(85) & 0.47518(69) \\
48   & 1.33504(93) & 0.39503(54) & 1.33506(93) & 0.39503(54) \\
40   & 1.34177(77) & 0.29577(34) & 1.34179(77) & 0.29577(34) \\
32   & 1.34958(71) & 0.20728(21) & 1.34960(71) & 0.20728(21) \\
27   & 1.35514(60) & 0.15838(14) & 1.35515(60) & 0.15838(14) \\
24   & 1.35836(53) & 0.13192(11) & 1.35836(53) & 0.13192(11) \\
20   & 1.36471(61) & 0.098821(75) & 1.36471(61) & 0.098821(75) \\
16   & 1.37072(45) & 0.069508(41) & 1.37072(45) & 0.069508(41) \\
\hline
\end{tabular}
\end{center}
\end{table}

\section{Estimation of the critical coupling}
\label{betacest}

\begin{figure}
\begin{center}
\includegraphics[scale=0.3]{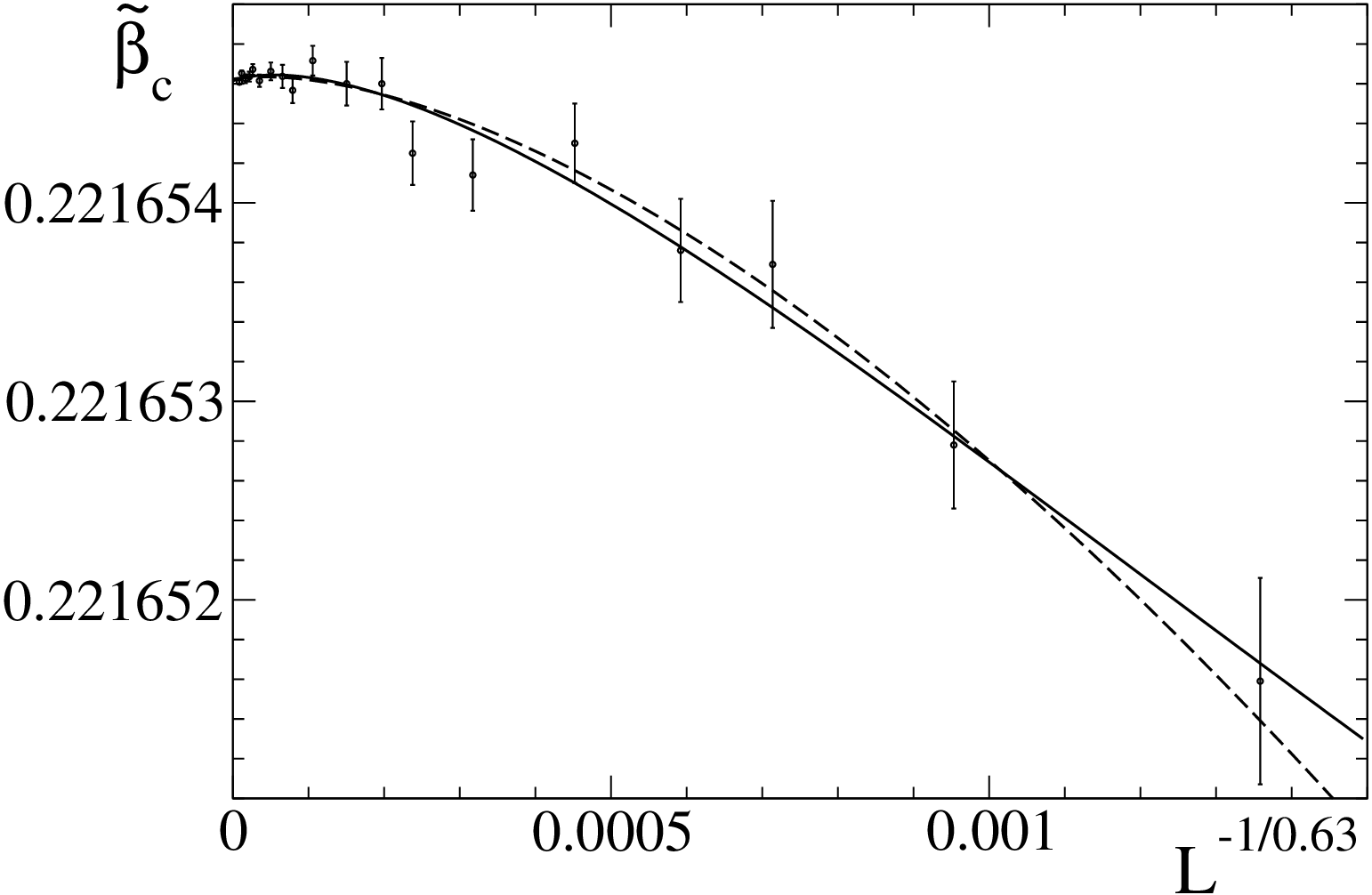}
\end{center}
\caption{The pseudo-critical coupling $\widetilde \beta_c$ vs 
$L^{-1/0.63}$. The dashed line shows the fit to~(\ref{betac}), including only the leading 
correction term to $\widetilde{\beta}_c - \beta_c$ with coefficient $a_1$ and exponents $\nu=0.63$ and $\omega=0.8$,
whereas the solid line is the fit with both correction terms included.}
\label{betac1}
\end{figure}

We have evaluated the critical coupling $\beta_c$ by fitting the $\widetilde{\beta_c}(L)$ data
to the finite--size scaling ansatz
\begin{equation}
\widetilde{\beta}_c(L) \simeq \beta_c + L^{-1/\nu} \left(a_0 + a_1 L^{-\omega} + a_2 L^{-2 \omega} \right) \;,
\label{betac}
\end{equation}
neglecting higher order corrections. The fits with the usual (RG) exponents $\nu=0.63$ and $\omega=0.8$
within the range $L \in [64,1536]$ are shown in Fig.~\ref{betac1}. 
The dashed line shows the fit with
only the leading correction term (the term with coefficient $a_1$) to $\widetilde{\beta_c} - \beta_c$ included, 
whereas the solid line -- the fit with both correction terms. These fits give $\beta_c = 0.2216546234(99)$ 
and $\beta_c = 0.221654615(13)$, respectively. The $\chi^2/\mathrm{d.o.f.}$ of both these fits is $1.08$. 
The data for $L < 64$ are omitted here, since inclusion of these relatively small sizes only increases the 
systematic errors without an essential reduction of the statistical errors. 

\begin{figure}
\begin{center}
\includegraphics[scale=0.3]{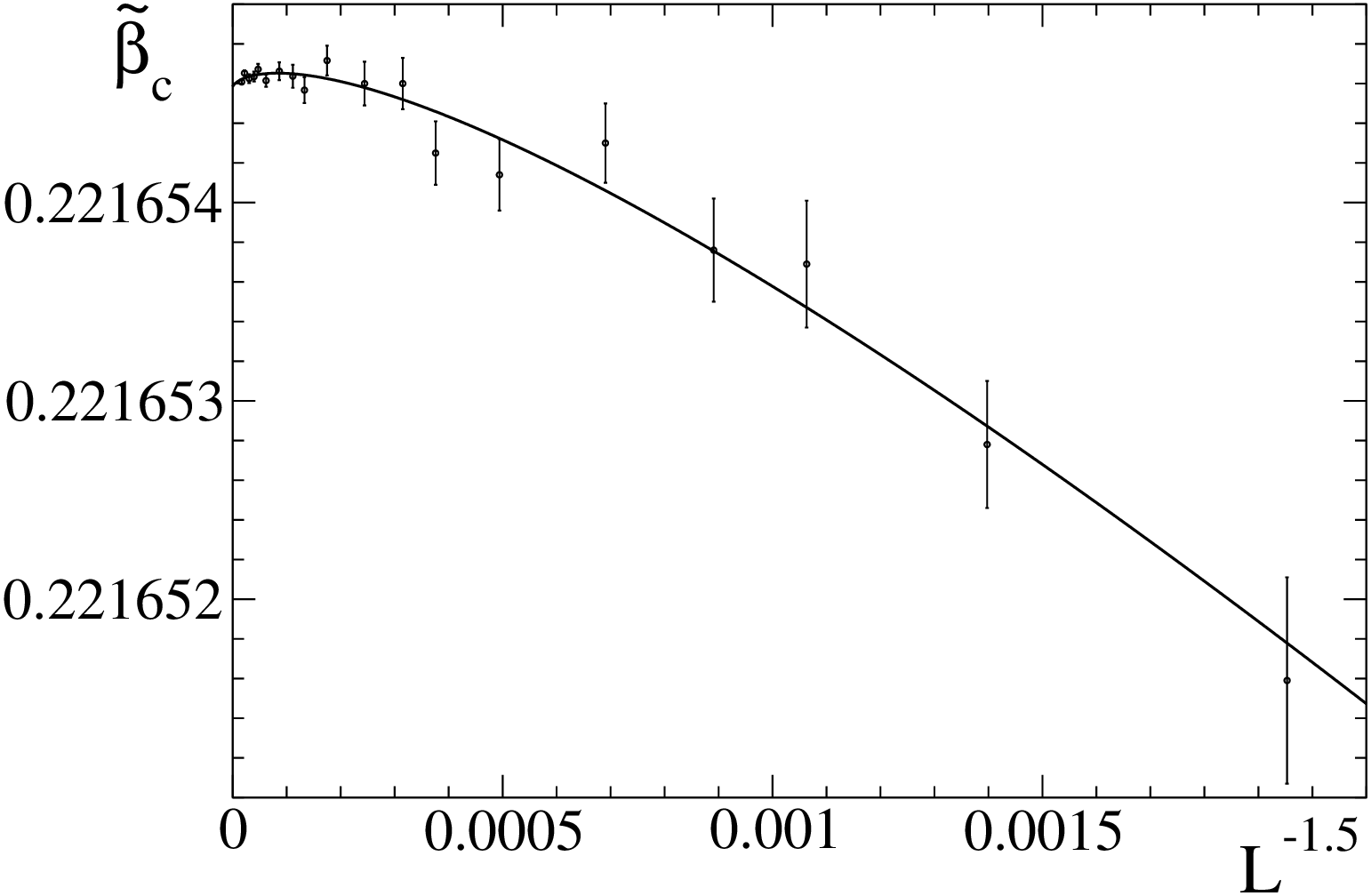}
\end{center}
\caption{The pseudo-critical coupling $\widetilde \beta_c$ vs 
$L^{-3/2}$. The solid curve shows the fit to~(\ref{betac}), including only the leading 
correction term to $\widetilde{\beta}_c - \beta_c$ with coefficient $a_1$ and exponents $\nu=2/3$ and $\omega=1/8$.}
\label{betac2}
\end{figure}

Surprisingly, the $\widetilde \beta_c$ data within $L \in [64,1536]$
can be even better fit with the exponents $\nu=2/3$ and $\omega=1/8$ of the GFD theory.
The fit, with only the leading correction to $\widetilde{\beta}_c - \beta_c$ included,
is shown in Fig.~\ref{betac2}. It yields $\beta_c = 0.221654585(15)$. In this case $\chi^2/\mathrm{d.o.f.}$
has the value $0.97$, which is even smaller than for the fits with $\nu=0.63$ and $\omega=0.8$.
Inclusion of the second correction term changes the result only slightly, i.~e., 
it gives $\beta_c = 0.221654588(47)$ with $\chi^2/\mathrm{d.o.f.}= 1.03$,
and the fit curve lies almost on the top of that one shown in Fig.~\ref{betac2}.
The statistical error, however, is strongly increased in this case. 

\begin{figure}
\begin{center}
\includegraphics[scale=0.3]{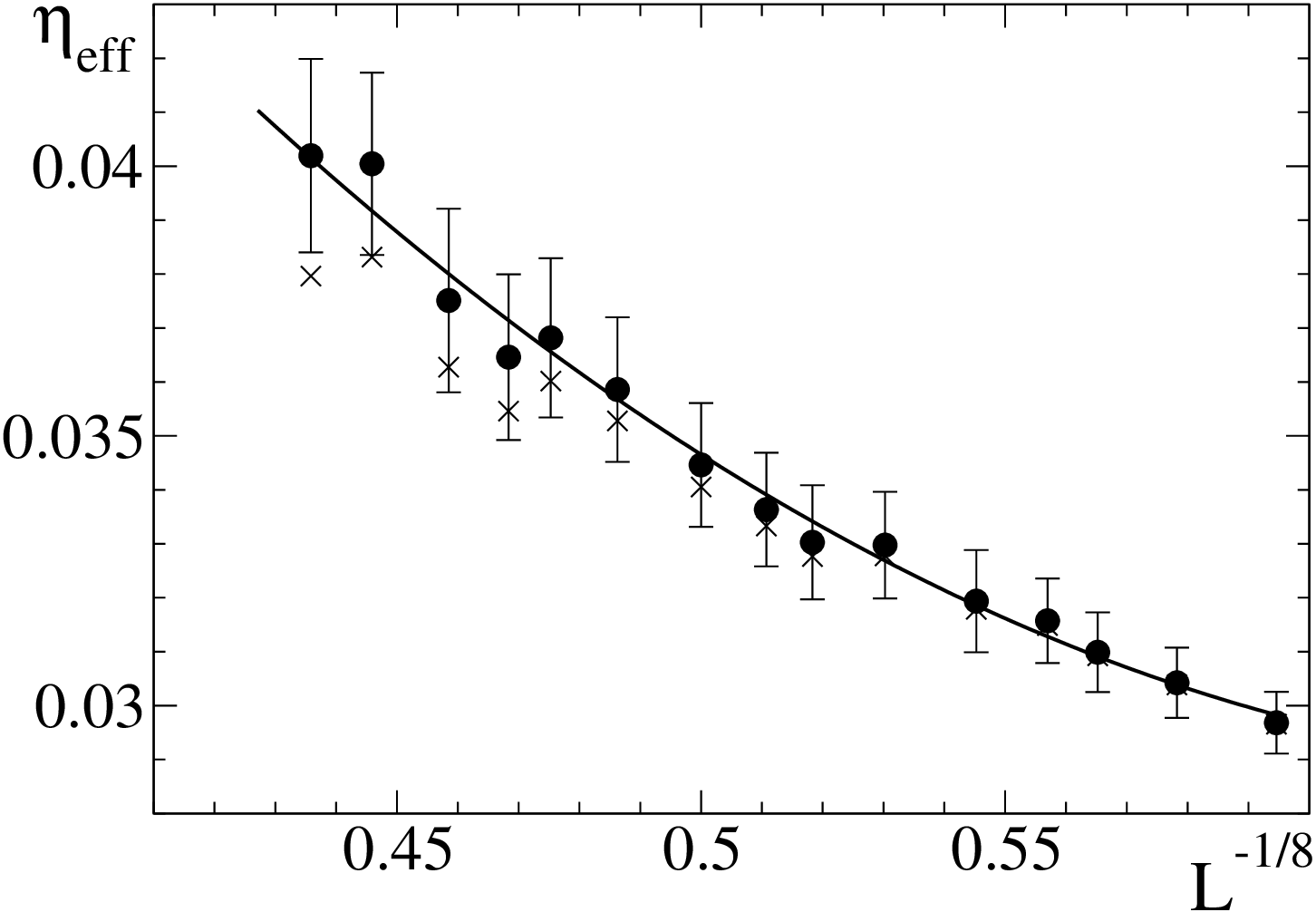}
\end{center}
\caption{The effective exponent 
$\eta_{\mathrm{eff}}$ vs $L^{-1/8}$ at $\beta=0.221654604$ (solid circles) and
$\beta=0.221654615$ (x). The solid line is the fit of circles to~(\ref{etaeff}) with $\eta=\omega=1/8$.}
\label{etam}
\end{figure}

Alternatively, the critical coupling $\beta_c$ can be determined from the susceptibility data,
requiring the consistency with the finite--size scaling ansatz
\begin{equation}
\chi \simeq L^{2-\eta} \left( b_0 + b_1 L^{-\omega} + b_2 L^{-2 \omega}  \right) 
\label{chi}
\end{equation}
or with the corresponding ansatz for the effective exponent 
\begin{equation}
\eta_{\mathrm{eff}}(L) \simeq \eta + c_1 L^{-\omega} + c_2 L^{-2 \omega}
\label{etaeff}
\end{equation}
at $\beta = \beta_c$. Here the effective exponent $\eta_{\mathrm{eff}}(L)$ is the local slope
of the $-\ln \left(\chi/L^2  \right)$ vs $\ln L$ plot, evaluated from the linear fit within $[L/2,2L]$.
We observe that the plot of the effective exponent within $L \in [64,768]$ (evaluated
from the data within $L \in [32,1536]$) is optimally 
described by~(\ref{etaeff}) with $\eta=\omega=1/8$ at $\beta \simeq 0.221654604$, as shown in Fig.~\ref{etam}.
From this we conclude that, if the exponents $\eta=\omega=1/8$ are correct, then
$\beta_c=0.221654604(18)$. This is the estimate of $\beta$ at which the fit to~(\ref{etaeff}) with
$\omega=1/8$ and $\eta$ as a fit parameter yields $\eta=1/8$ within the statistical
error bars. The solid--line fit at $\beta=0.221654604$ is really good, and the effective exponents
can be quite well fit with $\eta=\omega=1/8$ also at $\beta=0.221654615$ (the data shown by ``x''),
which is the best $\beta_c$ value provided by the already considered estimation with the RG exponents.

The value $\beta_c=0.221654604(18)$ is our best estimate of the critical coupling in the case where we use the exponents
of the GFD theory, since two corrections to scaling are included, giving smaller statistical error bars than for the
other estimate $\beta_c = 0.221654588(47)$ with two corrections included. Our estimated value 
$\beta_c=0.221654604(18)$ is sufficiently reasonable and plausible, since it agrees well within 
the error bars with all other estimations considered here, as well as with the most accurate values of $\beta_c$
provided by other authors, i.~e., $\beta_c=0.22165455(5)$~\cite{DBJ03} and $\beta_c=0.22165457(3)$~\cite{PFL02}.

Concerning the standard RG exponents, the method of fitting $\eta_{\mathrm{eff}}$ data is less convenient, 
since it is sensitive to the precise value of $\eta$, which is not known accurately enough in this approach.

\section{Finite--size scaling analysis of the critical exponents}

\subsection{Tests of the RG exponents}

In order to test the results of the perturbative RG theory, we have estimated the critical exponent 
$\eta$ by fitting the susceptibility data to~(\ref{chi}) and also by fitting the $\eta_{\mathrm{eff}}$ 
data to~(\ref{etaeff}) at fixed RG exponent $\omega=0.8$. Note that~(\ref{chi}) and~(\ref{etaeff})
are valid at $\beta=\beta_c$, as well as at $\beta=\widetilde{\beta}_c(L)$. The fits have been made at
$\beta=\widetilde{\beta}_c(L)$ and at the estimated in Sec.~\ref{betacest} values of the critical coupling, 
obtained by using the RG exponents. 

\begin{figure}
\begin{center}
\includegraphics[scale=0.3]{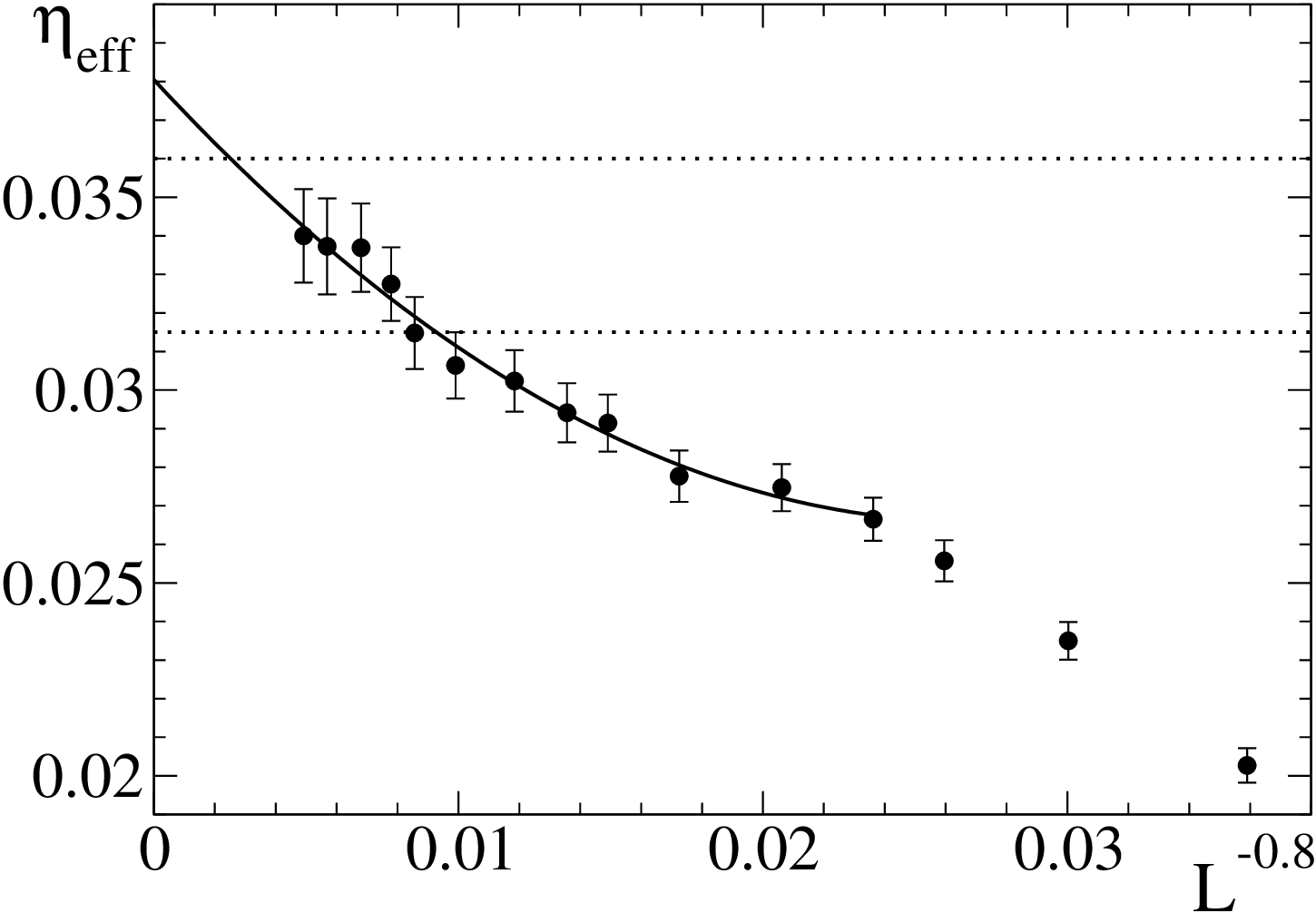}
\end{center}
\caption{The effective exponent 
$\eta_{\mathrm{eff}}$ vs $L^{-0.8}$ at $\beta=\widetilde{\beta}_c(L)$.
The solid curve is the fit to~(\ref{etaeff}) with fixed $\omega=0.8$.
The dotted lines indicate the range $[0.0315,0.036]$ of $\eta$
values consistent with the set of perturbative RG estimations in~\cite{GJ98}.}
\label{etap}
\end{figure}

\begin{figure}
\begin{center}
\includegraphics[scale=0.3]{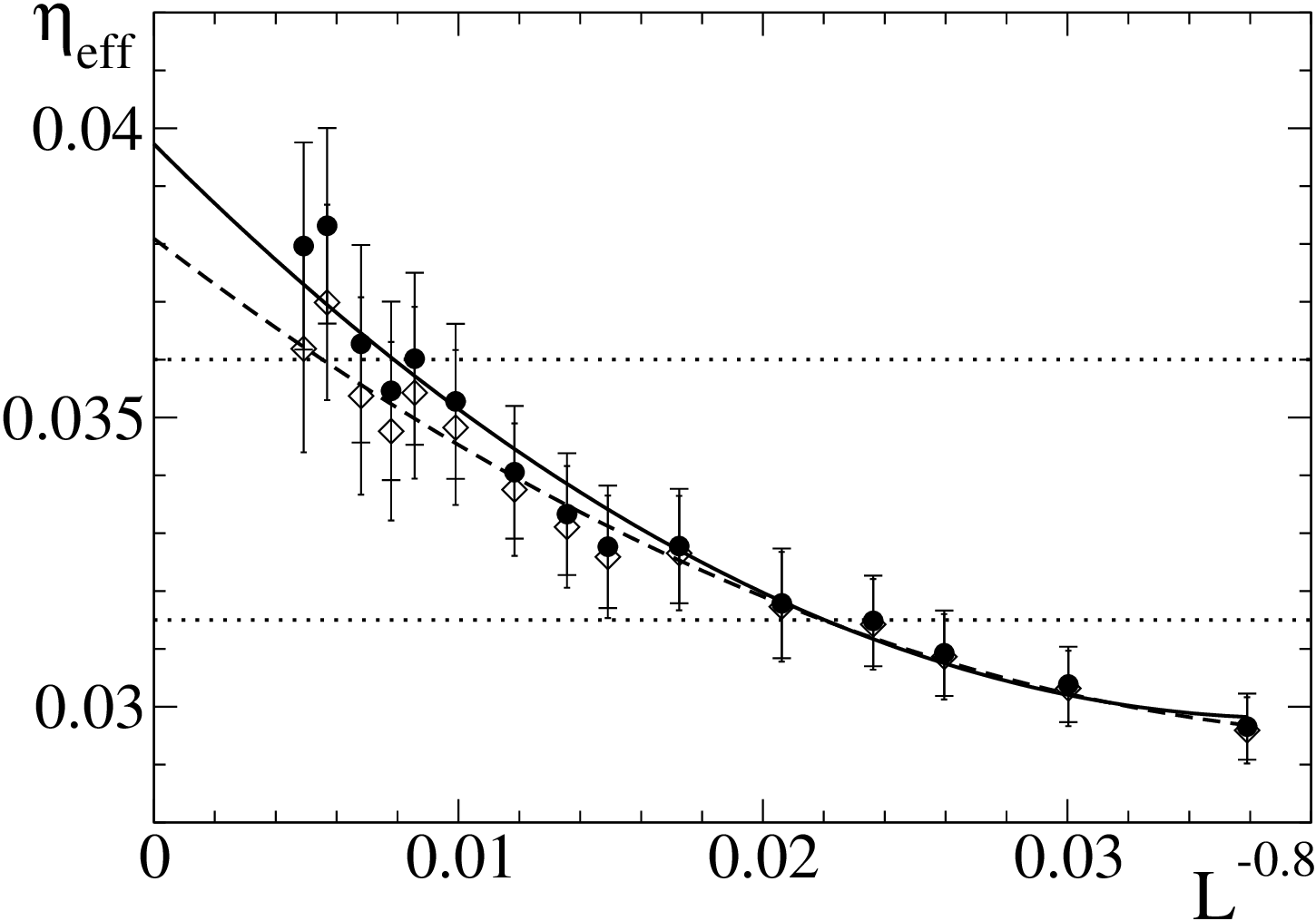}
\end{center}
\caption{The effective exponent 
$\eta_{\mathrm{eff}}$ vs $L^{-0.8}$ at $\beta_c=0.2216546234(99)$ (diamonds)
and $\beta_c=0.221654615(13)$ (circles). The curves are the 
fits to~(\ref{etaeff}) with fixed $\omega=0.8$.
The dotted lines indicate the range $[0.0315,0.036]$ of $\eta$
values consistent with the set of perturbative RG estimations in~\cite{GJ98}.}
\label{eta}
\end{figure}

We start with the fitting of the effective exponent, since this method can be better controlled to see
which fits are most appropriate. The plot of $\eta_{\mathrm{eff}}$ vs $L^{-\omega}$ with $\omega=0.8$, 
evaluated at $\beta=\widetilde{\beta}_c(L)$, is shown in Fig.~\ref{etap}
We see that the data are well described by a quadratic curve for $L \ge 108$,
therefore one has to include both correction terms in~(\ref{etaeff}). The fit
yields $\eta=0.0380(26)$. This value is slightly larger than the upper limit
$0.036$ of the interval $0.0315 < \eta < 0.036$ for the values, which are consistent 
with the set of perturbative RG estimations of~\cite{GJ98} discussed in Sec.~\ref{cps}.
However, the discrepancy is within the error bars.
Somewhat larger deviations above $0.036$ are provided by the fits within $L \ge 64$, 
shown in Fig.~\ref{eta}, at approximately estimated values $\beta_c=0.2216546234(99)$ 
and $\beta_c=0.221654615(13)$ of the critical coupling.
These fits yield $\eta = 0.0381(18)$ and $\eta=0.0397(28)$, respectively.
The indicated here standard errors are calculated by the jackknife method, 
taking into account the statistical fluctuations in the estimated $\beta_c$ value.
Namely, the standard error $\sigma$ is evaluated as $\sigma = \sqrt{\sum_{ij} \Delta_{ij}^2}$,
where $\Delta_{ij}$ is the shift in the estimated $\eta$ value, taking into account
also the shift in the fitted $\beta_c$ value, when the $j$th simulation block
(iteration) for the $i$th lattice size is omitted. The statistical correlations are such 
that the total standard error is smaller than in the case where there are no correlations 
between the fluctuations in the estimated $\beta_c$ value and in the $\eta$ value obtained 
at a given $\beta$. 

We have also evaluated $\eta$ from the fits to~(\ref{chi}) at the pseudo-critical coupling 
$\widetilde{\beta}_c(L)$, as well as at the two estimates
$\beta_c=0.2216546234(99)$ and $\beta_c=0.221654615(13)$ of the critical coupling.
The results of fits within $L \in [L_{\mathrm{min}},1536]$, assuming $\omega=0.8$, are 
collected in Tab.~\ref{tab3}.
\begin{table}
\caption{The critical exponent 
$\eta$ evaluated from the fits to~(\ref{chi}) within $L \in [L_{\mathrm{min}},1536]$
with fixed $\omega=0.8$. The estimates from the data at $\beta=\widetilde{\beta}_c(L)$ and
at $\beta=\beta_c$ with $\beta_c=0.2216546234(99)$ and $\beta_c=0.221654615(13)$ 
are denoted by $\eta_1$, $\eta_2$ and $\eta_3$, respectively.}
\label{tab3}
\begin{center}
\begin{tabular}{|c|c|c|c|}
\hline
\rule[-2mm]{0mm}{7mm} 
 $L_{\mathrm{min}}$  & $\eta_1$ & $\eta_2$  & $\eta_3$  \\
\hline
32 & 0.0345(12) & 0.0363(14) & 0.0379(25)  \\
40 & 0.0353(14) & 0.0364(15) & 0.0384(27)  \\
48 & 0.0359(17) & 0.0369(17) & 0.0394(29)  \\
54 & 0.0369(19) & 0.0376(20) & 0.0402(28)  \\
64 & 0.0380(23) & 0.0376(30) & 0.0405(25)  \\
80 & 0.0377(37) & 0.0366(38) & 0.0398(32)  \\
\hline
\end{tabular}
\end{center}
\end{table}
The best estimates in these three cases are assumed to be
$\eta=0.0380(23)$, $\eta=0.0376(20)$ and $\eta=0.0405(25)$, since they perfectly agree 
with the corresponding estimates at a larger $L_{\mathrm{min}}$, but have smaller
statistical errors. Note that the decrease of the standard error for $\eta_3$ in Tab.~\ref{tab3}
at $L_{\mathrm{min}}=64$ as compared to $L_{\mathrm{min}}=48, 54$ is the effect
of the already mentioned statistical correlations.
These best $\eta$ values are very similar to the ones 
obtained before from the effective exponents.

Remarkable is the fact that a better estimation of the critical coupling,
including two corrections to scaling instead of only one correction,
leads to a worse agreement with the results of the perturbative RG theory.
In particular, the value $\eta=0.0405(25)$ deviates above $0.036$
(the upper limit of the best RG estimate $\eta=0.0335 \pm 0.0025$ of~\cite{GJ98})
by $1.8 \sigma$. The observed here deviations
suggest that $\eta$, probably, is larger than normally expected 
from the perturbative RG theory. The statistical errors, however,
do not allow to make a strict conclusion.

\subsection{Fits with the exponents of the GFD theory}
\label{sec:our}

\begin{figure}
\begin{center}
\includegraphics[scale=0.3]{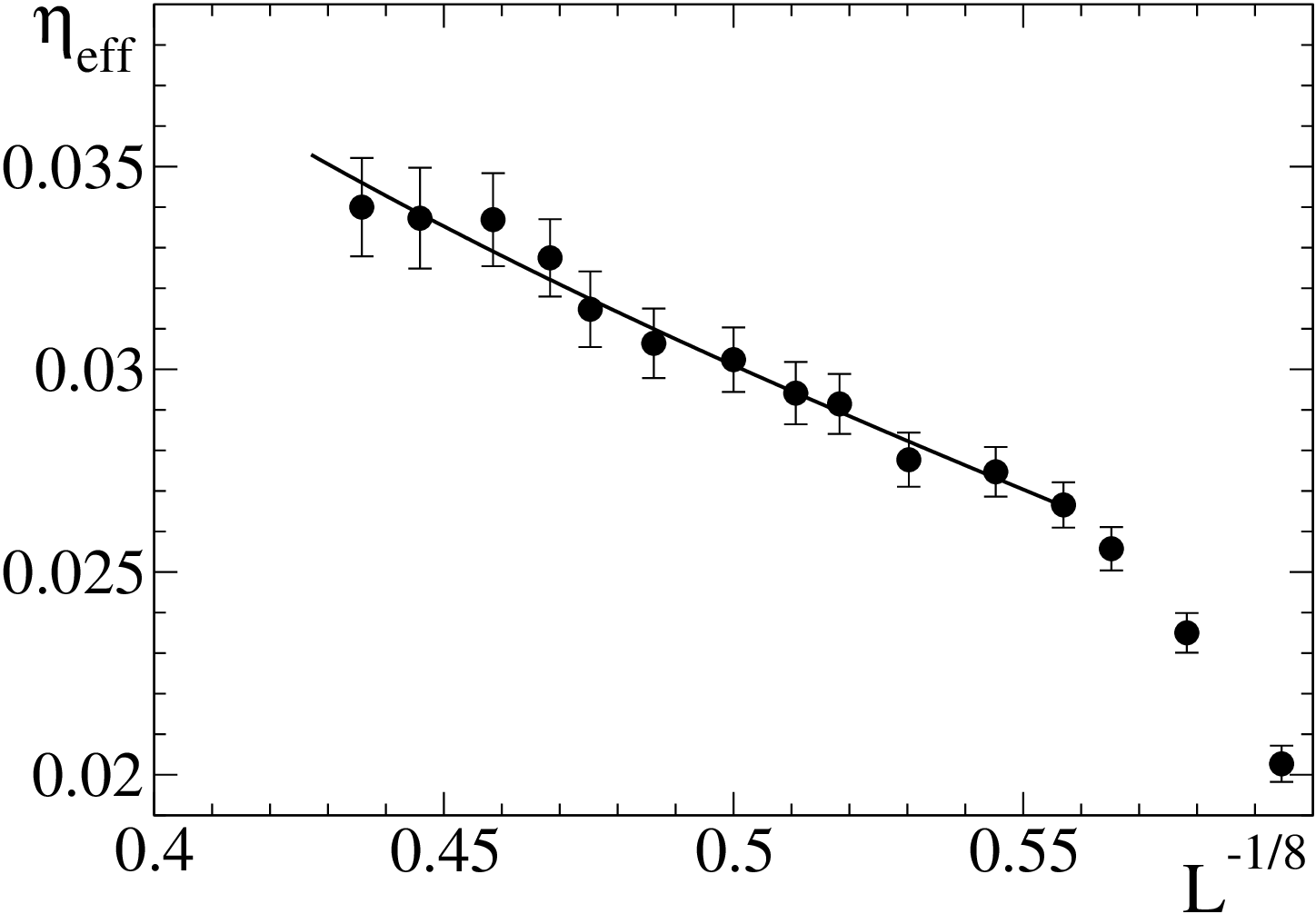}
\end{center}
\caption{The effective exponent 
$\eta_{\mathrm{eff}}$ vs $L^{-1/8}$ at $\beta=\widetilde{\beta}_c(L)$.
The solid line is the fit with $\eta=\omega=1/8$.}
\label{etapm}
\end{figure}

As already discussed in Sec.~\ref{betacest}, the pseudo-critical coupling 
$\widetilde{\beta}_c(L)$ can be even better fit with the exponents $\nu=2/3$ and $\omega=1/8$
of the GFD theory than with those of the perturbative RG theory. We have also verified 
that the effective exponent $\eta_{\mathrm{eff}}(L)$ is very well described by
the ansatz~(\ref{etaeff}) with $\eta=\omega=1/8$ (see Fig.~\ref{etam}) at a certain $\beta$ corresponding to a reasonable 
estimate of the critical coupling $\beta_c=0.221654604(18)$. Also at the pseudo-critical coupling
$\beta=\widetilde{\beta}_c(L)$, the effective exponent $\eta_{\mathrm{eff}}(L)$ within $L \in [108,768]$
(extracted from the susceptibility data within $L \in [54,1536]$) can be well fit to~(\ref{etaeff})
with fixed GFD exponents $\eta=\omega=1/8$, as shown in Fig.~\ref{etapm}.
If the exponent $\eta$ is considered as a fit parameter at fixed $\omega=1/8$, then
this fit yields $\eta=0.079(80)$ in agreement with the expected value $1/8$, although
the statistical error is rather large in this case.  Sufficiently accurate data
for even larger $L$ values would be very helpful here to perform a more precise testing 
of the agreement. 

We have verified that not only the effective exponent $\eta_{\mathrm{eff}}$, but also
the susceptibility data can be well described by the exponents $\eta=\omega=1/8$.
In particular, $\chi^2/\mathrm{d.o.f}$ of the fit with $\eta=\omega=1/8$ is $0.88$ for the susceptibility data
 at $\beta=\widetilde{\beta}_c(L)$ within $L \in [64,1536]$. At $\beta=0.221654604 \simeq \beta_c$,
the data can be well fit with $\chi^2/\mathrm{d.o.f}=1.01$ over a wider range of sizes $L \in [48,1536]$.

We have fitted also the $\partial Q/ \partial \beta$ data to the finite--size scaling ansatz
\begin{equation}
\frac{\partial Q}{\partial \beta} \simeq L^{1/\nu} \left( A_0 + A_1 L^{-\omega} + A_2 L^{-2 \omega} \right) \;.
\end{equation}
These data can be well fit with the RG exponents, as well as with those of the GFD theory.
In the latter case ($\nu =2/3$, $\omega=1/8$), the data within $L \in [32,1536]$
are fit  with $\chi^2/\mathrm{d.o.f}=0.88$ at $\beta=\widetilde{\beta}_c(L)$ and
with $\chi^2/\mathrm{d.o.f}=0.93$ at $\beta=0.221654604 \simeq \beta_c$.

In fact, reasonable fits with the exponents $\eta=1/8$ and $\nu=2/3$ are possible 
because of the value $1/8$ of the correction--to--scaling exponent $\omega$.
A remarkably larger value, such as $\omega \simeq 0.8$, would make such fits not good.
According to~(\ref{etaeff}), the plot of the effective exponent $\eta_{\mathrm{eff}}$ vs 
$L^{-\omega}$ has to be almost linear at $L \to \infty$. If $\omega$ is as large as
$0.8$, then $L^{-2 \omega}$ is already quite small for $L \ge 108$, so that a good linearity
of the fit curve in Fig.~\ref{etap} can be expected. However, the plot looks rather
nonlinear. Moreover, it can be much better approximated by a straight line (within $L \in [108,768]$)
in the $L^{-1/8}$ scale than in the $L^{-0.8}$ scale, as it is clearly seen comparing
the plots in Figs.~\ref{etap} and~\ref{etapm}. According to this, it seems, in fact, very plausible
that $\omega$ could be about $1/8$.

\section{Conclusions}

The analysis of the MC data for the $O(2)$ and $O(4)$ models below the critical
point supports the recently published theoretical results~\cite{K2010},
predicting that the Goldstone mode singularities in the $O(n)$ 
models are described by nontrivial exponents, as discussed in Sec.~\ref{sec:Goldstone}.
Therefore, it has been important to verify the earlier predictions
of this approach, called the GFD theory, concerning the critical point
singularities. For this purpose, we have performed MC simulations of the 3D Ising model
for very large lattices with linear size up to $L=1536$, using a parallel implementation 
of the Wolff single cluster algorithm. The finite--size scaling analysis shows
that the actual data can reasonably well interpreted with the usual
critical exponents $\eta \simeq 0.0335$, $\nu \simeq 0.63$ and $\omega \simeq 0.8$
of the perturbative RG theory, and can also be well described
by a different set of critical exponents, $\eta=\omega=1/8$ and $\nu=2/3$,
found within the GFD theory. The validity of the fits with the latter set of 
exponents depends on whether the correction--to--scaling exponent $\omega$ 
can, indeed, be as small as $1/8$. We have demonstrated in Sec.~\ref{sec:our} 
that this is a very plausible scenario.

We have performed a high accuracy estimation of the critical
coupling $\beta_c$, combined with a sophisticated fitting of the data,
including the leading, as well as sub-leading, corrections to scaling, to distinguish
between the two sets of the critical exponents.  
We have found that a self-consistent estimation of $\beta_c$ and $\eta$, assuming the RG exponents $\nu \simeq 0.63$
and $\omega=0.8$, gives somewhat larger values of $\eta$ (e.~g., $\eta=0.0397(28)$ 
and $\eta=0.0405(25)$ when two corrections to scaling are included) than the value $0.0335 \pm 0.0025$ 
expected from the most accurate resummation of the RG perturbative series~\cite{GJ98}. 
These discrepancies are not large enough to make a strict conclusion about an inconsistency.
On the other hand, the observed gradual increase in the effective exponent $\eta_{\mathrm{eff}}(L)$ with growing
system size $L$ is just expected if $\eta=\omega=1/8$ and $\nu=2/3$ are the correct asymptotic exponents. 
Therefore, it would be very interesting and important to see whether such an inrease to even larger 
$\eta_{\mathrm{eff}}$ values is supported by the data for even larger lattice sizes.

\section*{Acknowledgments}

This work was made possible by the facilities of the
Shared Hierarchical Academic Research Computing Network
(SHARCNET:www.sharcnet.ca). R. M. acknowledges the support from the
NSERC and CRC program.
We thank also the DEISA Consortium (www.deisa.eu), funded through the EU FP7
project RI-222919, for support within the DEISA Extreme Computing Initiative.

\end{document}